\documentclass[amsmath,onecolumn,superscriptaddress,notitlepage]{revtex4-1}
\usepackage[normalem]{ulem}
\usepackage{graphicx}

\begin{document}

\title{High-fidelity phase and amplitude control of phase-only computer generated holograms using conjugate gradient minimisation}

\author{D.~Bowman}
\affiliation{SUPA School of Physics \& Astronomy, University of St.~Andrews, North Haugh, St.~Andrews KY16 9SS, UK}
\author{T.~L.~Harte}
\affiliation{Clarendon Laboratory, University of Oxford, Parks Road, Oxford OX1 3PU, UK}
\author{V.~Chardonnet}
\affiliation{SUPA School of Physics \& Astronomy, University of St.~Andrews, North Haugh, St.~Andrews KY16 9SS, UK}
\affiliation{Sorbonne Universit\'{e}s, UPMC Univ Paris 06, UFR 925, 4 place Jussieu, 75252 Paris cedex 05, France}
\author{C.~De~Groot}
\affiliation{SUPA School of Physics \& Astronomy, University of St.~Andrews, North Haugh, St.~Andrews KY16 9SS, UK}
\author{S.~J.~Denny}
\affiliation{Clarendon Laboratory, University of Oxford, Parks Road, Oxford OX1 3PU, UK}
\author{G.~Le~Goc}
\affiliation{SUPA School of Physics \& Astronomy, University of St.~Andrews, North Haugh, St.~Andrews KY16 9SS, UK}
\affiliation{Sorbonne Universit\'{e}s, UPMC Univ Paris 06, UFR 925, 4 place Jussieu, 75252 Paris cedex 05, France}
\author{M.~Anderson}
\affiliation{SUPA School of Physics \& Astronomy, University of St.~Andrews, North Haugh, St.~Andrews KY16 9SS, UK}
\author{P.~Ireland}
\affiliation{SUPA School of Physics \& Astronomy, University of St.~Andrews, North Haugh, St.~Andrews KY16 9SS, UK}
\author{D.~Cassettari}
\affiliation{SUPA School of Physics \& Astronomy, University of St.~Andrews, North Haugh, St.~Andrews KY16 9SS, UK}
\author{G.~D.~Bruce}
\affiliation{SUPA School of Physics \& Astronomy, University of St.~Andrews, North Haugh, St.~Andrews KY16 9SS, UK}
\email{gdb2@st-andrews.ac.uk} %% email address is required



\begin{abstract}
We demonstrate simultaneous control of both the phase and amplitude of light using a conjugate gradient minimisation-based hologram calculation technique and a single phase-only spatial light modulator (SLM). A cost function which incorporates the inner product of the light field with a chosen target field within a defined measure region is efficiently minimised to create high fidelity patterns in the Fourier plane of the SLM. A fidelity of $F=0.999997$ is achieved for a pattern resembling an $LG^{0}_{1}$ mode with a calculated light-usage efficiency of $41.5\%$. Possible applications of our method in optical trapping and ultracold atoms are presented and we show uncorrected experimental realisation of our patterns with $F = 0.97$ and $7.8\%$ light efficiency. %\textcolor{red}{SHOULD BE APPROX. 100 WORDS}
\end{abstract}

\maketitle

%\ocis{ (050.1970) Diffractive Optics; (090.1760) Computer holography; (090.1995) Digital holography;
%(230.6120) Spatial light modulators; (020.7010) Laser trapping}
%For a complete list of OCIS codes, visit: https://www.osapublishing.org/oe/submit/ocis/

%%%%%%%%%%%%%%%%%%%%%%%%%%  body  %%%%%%%%%%%%%%%%%%%%%%%%%%
\section{Introduction}

Simultaneous control over the amplitude and phase of light has allowed significant advances in optical trapping of microscopic objects \cite{Woerdemann+13}, microscopy \cite{Maurer+11} and optical communication \cite{Willner+15}. A variety of methods have been developed which allow arbitrary independent control over both. Tandem or cascaded approaches sequentially manipulate the amplitude then phase using either two Spatial Light Modulators (SLMs) or two distinct regions of a single SLM \cite{Neto+96,Jesacher+08,Zhu+14}. Analytical approaches which calculate a single phase-only modulation to simultaneously sculpt amplitude and phase include the shape-phase method \cite{Roichman+06} and a variety of methods which spatially control the height, and thus diffraction efficiency, of the applied phase \cite{Clark+16}. Recently, a high-fidelity superpixel approach to phase and amplitude control has also been demonstrated for Digital Micromirror Devices (DMDs) \cite{Goorden+14}.

In order to control the light field in a particular plane holographically, we wish to apply a bespoke phase modulation $\phi_{p,q}$ (with indices $p$ and $q$ denoting spatial co-ordinates) to a fixed incident laser field with amplitude $S_{p,q}$, in a simple setup with a single phase-only SLM and a single focussing element. The electric field in the plane of the SLM is $E^{\text{in}}_{p,q} = S_{p,q}\exp\left(i\phi_{p,q}\right)$. Given $S_{p,q}$ and $\phi_{p,q}$, the electric field in any other plane $E^{\text{out}}_{n,m}$ (with output plane coordinates denoted by $n$ and $m$) is straightforwardly calculated using an appropriate propagator $\mathcal{P}$ such that $E^{\text{out}}_{n,m}=\mathcal{P}\left[E^{\text{in}}_{p,q}\right]$. For patterns in the far field $\mathcal{P}$ is approximated by a fast Fourier transform \cite{Goodman+96} such that
\begin{align}
E^{\text{out}}_{n,m} &= \frac{S_{p,q}}{N_{T}}\sum_{p,q}\exp\left(i\phi_{p,q}\right)\exp\left[-\left(\frac{2\pi i}{N_{T}}\right)\left(pn + qm\right)\right], \label{equation:Eout1} \\
&= \sqrt{I_{n,m}}\exp\left(i\varphi_{n,m}\right), \label{equation:Eout2}
\end{align}
where $N_{T}=\sum_{n,m}1$, while $I_{n,m}$ and $\varphi_{n,m}$ are the output plane intensity and phase respectively. Calculation of the appropriate phase-only modulation $\phi_{p,q}$ to give an acceptable output field is a well-known inverse problem which, in general, requires numerical solution. Iterative Fourier Transform Algorithms (IFTAs) are commonly used in calculating the phase modulation required to generate a desired intensity distribution, and variants which control both phase and amplitude have been recently demonstrated \cite{Tao+15,Wu+15}.  

In this paper we propose an alternative iterative method to creating patterns with independent control over the phase and amplitude profiles: using a conjugate gradient minimisation technique which was previously shown to achieve smooth, accurate and highly-controllable intensity patterns \cite{Harte+14}. The technique efficiently minimises a specified cost function which can be carefully manipulated to reflect the requirements of the chosen light pattern, such as removing optical vortices from regions of interest. Here, we extend this method to produce a variety of high fidelity and smooth patterns in both phase and intensity, which are designed primarily for optical trapping. 

\section{Conjugate Gradient Method}

The conjugate gradient minimisation method is intuitively described in \cite{Shewchuck+94}, and our original conjugate gradient optimisation routine for control of the amplitude in holograms is presented in more detail in \cite{Harte+14}. %In brief, the method efficiently scans a parameter space in order to minimise a cost function $C$. 
The main advantage of this approach is the high level of control it gives over any feature of interest in the output plane, provided that the feature can be encapsulated within an analytical cost function $C$. This defines an effective error to be minimised, and judicious choice of the cost function terms can allow precise guiding of the hologram optimisation process.  %The flexibility inherent in cost function choice has allowed us to extend this method to achieve precise control not only over the output plane intensity but also its phase. 
For our holograms, the cost function is based on the difference between the calculated electric field and a chosen target, and the parameter space for the optimisation encompasses all the different phase distributions that the SLM can generate. In order to find a hologram which gives acceptable amplitude and phase, we find that a good choice is
\begin{align}
C &= 10^{d}\left(1 - \sum_{n,m} \text{Re}\left\lbrace\left| \tilde{\tau}_{n,m}^{*} \tilde{E}^{\text{out}}_{n,m} \right|\right\rbrace\right)^{2}, \label{equation:cost1_1} \\
&= 10^{d}\left(1 - \sum_{n, m} \sqrt{\tilde{I}_{n,m}\tilde{T}_{n,m}}\cos\left(\Phi_{n,m} - \varphi_{n,m}\right)\right)^{2},
\label{equation:cost1_2}
\end{align}
where $\tau_{n,m}=\sqrt{T_{n,m}}\exp\left(i\Phi_{n,m}\right)$ is the target electric field, and the over-tilde denotes normalisation over a specified region of interest, which is small compared to the total output plane. Similar to the MRAF method \cite{Pasienski+08}, we choose this region of interest to encompass regions of non-zero amplitude in the target pattern (known as the measure region) plus a surrounding area of zero intensity. Experimentally, the light which the algorithm places outside the region of interest can be spatially filtered. The multiplicative prefactor $10^{d}$ is used to increase the steepness of the cost function within the parameter space to improve convergence time and accuracy. 

\begin{figure}[ht!]
\begin{center}
\includegraphics[width = 0.7\linewidth]{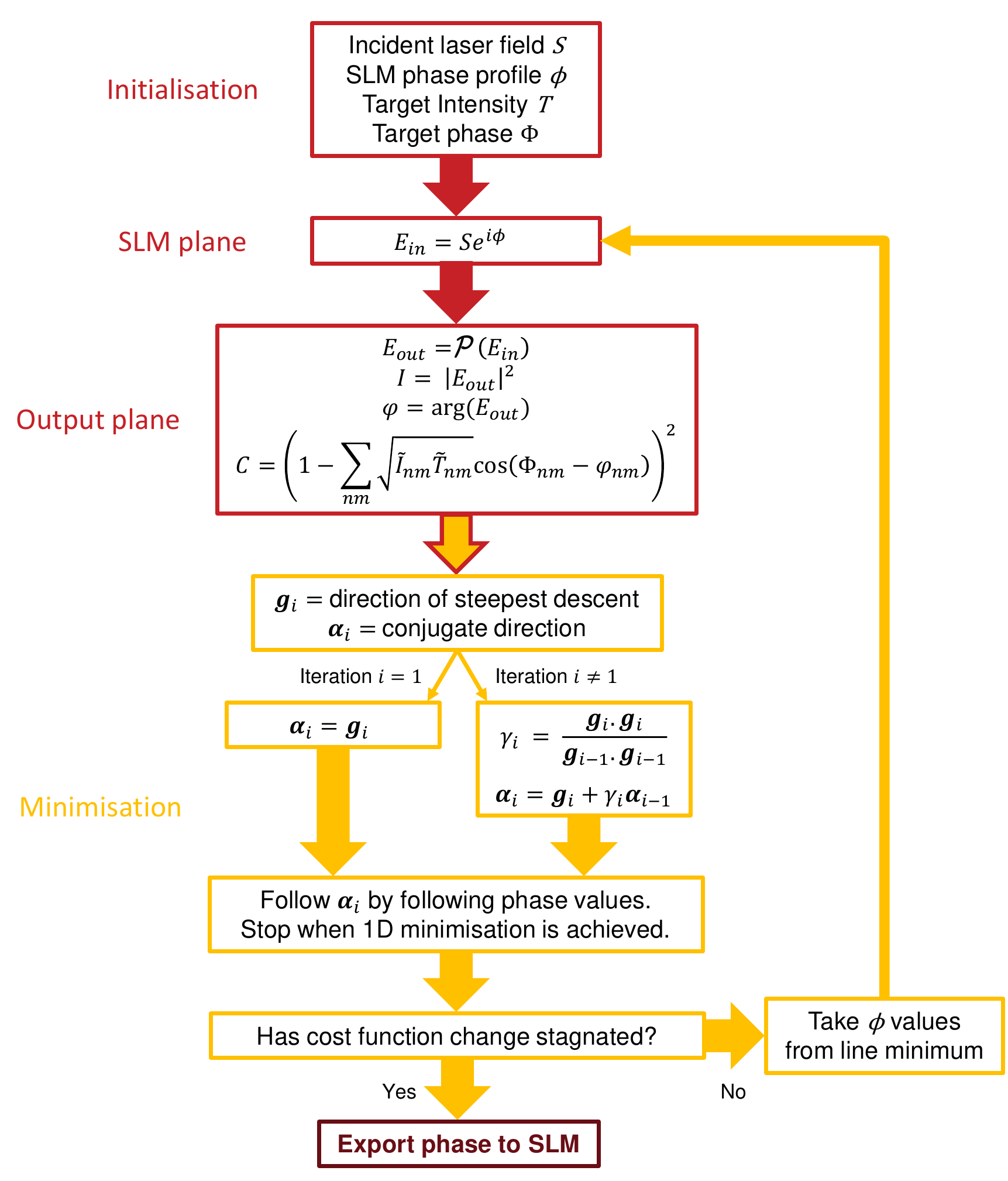}
\caption{Block diagram of the phase distribution calculation process using conjugate gradient minimisation. \label{fig:block}}
\end{center}
\end{figure}

Figure \ref{fig:block} shows a diagrammatic representation of the calculation. The initial position in the parameter space of $C$ is determined by $S_{p,q}$ (a two-dimensional Gaussian profile with $1/e^2$-radius $\sigma$) and a guess phase $\phi_{p,q}= R\left(p^2+q^2\right)+D \left(p \cos\theta+q\sin\theta\right)$. The two terms in $\phi_{p,q}$ respectively control the size and position of the envelope of the output plane intensity. This combination of phase patterns is known to suppress the formation of optical vortices during hologram calculation, which can otherwise cause premature stagnation and low accuracy \cite{Senthilkumaran+05,Pasienski+08}.

As an initial step, we calculate $\partial C / \partial \phi_{p,q}$ for each pixel to determine the direction of steepest descent $g_{1}$ and minimise $C$ along this direction to change $\phi_{p,q}$. For subsequent iterations $i$ of the process, the descent direction $\alpha_{i}$ is the conjugate direction
\begin{equation}
\alpha_{i}= g_{i} + \left(\frac{g_{i}.g_{i}}{\left(g_{i-1}.g_{i-1}\right)}\right)\alpha_{i-1}.
\label{equation:direction}
\end{equation}
The process continues until the cost function stagnates (i.e. when the difference in the value of the cost function between iterations is below $10^{-5}$) or a predefined maximum number of iterations is reached. We implement the conjugate gradient calculation in Python with the cost function gradient determined using the Theano library \cite{Theano}. %\textcolor{red}{Our codes, and the data presented in this article, are freely available online. \cite{OpenData}.}

\section{Numerical Results}

\begin{figure}[ht!]
\begin{center}
\includegraphics[width = 0.6\linewidth]{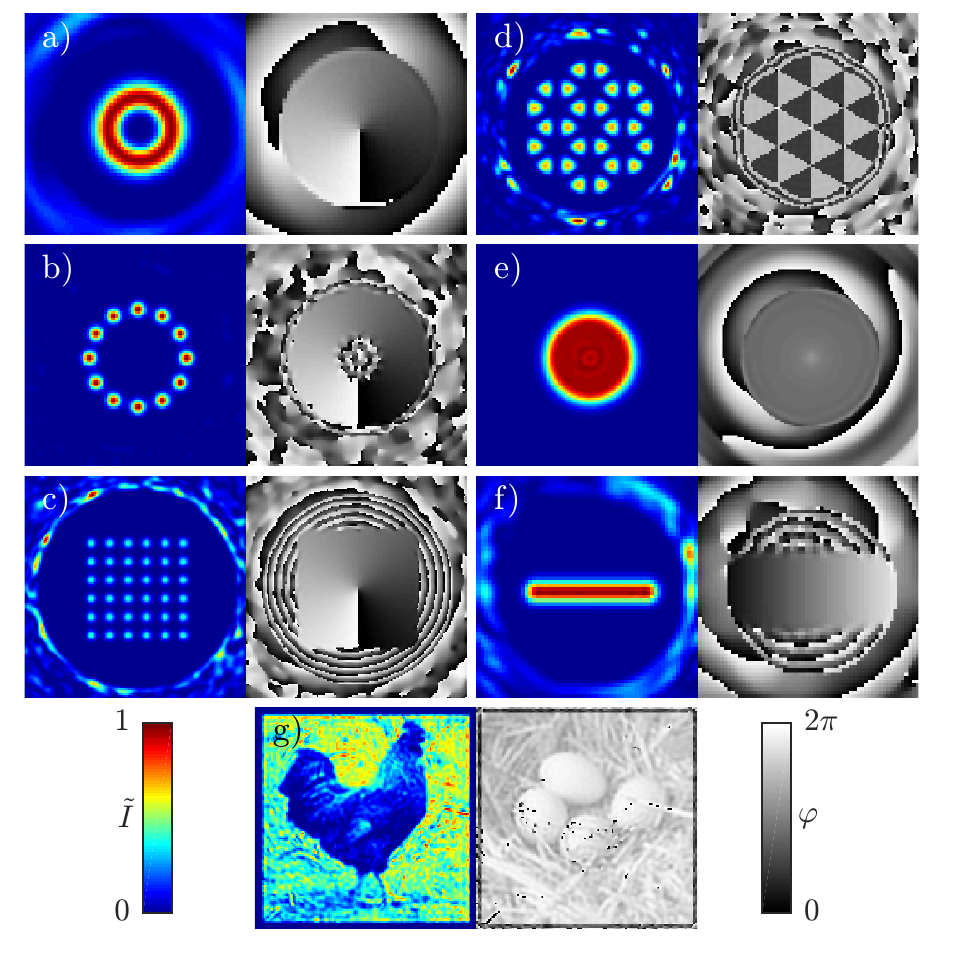}
\caption{The far-field results from the conjugate gradient optimisation showing normalised intensity $\tilde{I}$ (colour) and phase $\varphi$ (grey) in the region of interest. The flat top pattern (e) has the light outside the measure region removed for clarity. The error metrics for each pattern are shown in Table \ref{tabular:results1}. \label{fig:results1}}
\end{center}
\end{figure}

We test our method on a range of target patterns particularly chosen with applications in optical trapping in mind. Independent spatial control over both the amplitude and phase of trap light is also increasingly desirable in the field of ultracold atoms, for example in the transfer of orbital angular momentum from light to atoms \cite{Ramanathan+11}, and in the creation of artificial gauge fields \cite{Huo+14,Lembessis+15,Butera+16}. In the particular case of trapping ultracold atoms in continuous geometries \cite{Pasienski+08,Bruce+11,Gaunt+12,Bowman+15,Bruce+15,Buccheri+16}, accuracy and smoothness of the intensity are vital to avoid fragmentation. 

We calculate a pattern of phase values between 0 and $2\pi$ for the SLM plane of $256\times 256$ pixels (with a pixel size of $24\mu\text{m}$) padded with zeros in the border such that the plane is $512\times 512$ pixels, such that there is no loss of resolution in the resulting $512\times 512$ output plane. The patterns are diagonally offset from the center of the plane by 85 pixels to avoid the zeroth order (undiffracted light) that would appear due to the finite efficiency of the SLM. This constrains two of the initialisation parameters to $D = -\pi/2$ and $\theta=\pi/4$. 

We show the region of interest of the calculated intensity and phase for each of our target patterns in Figure \ref{fig:results1}. The pattern similar to a Laguerre-Gaussian (LG) mode provides a good benchmark for our method and such patterns have a wide variety of uses \cite{Yao+11}, including in ultracold atom experiments to induce circulation states \cite{Ramanathan+11}. We can also retain the phase structure of LG modes but with arbitrary amplitude profiles. As examples, ring and square lattices with underlying phase windings have potential applications for quantum simulation of magnetic flux in solid state systems \cite{Huo+14}.  Ultracold atoms confined in a honeycomb lattice with alternating phase between nearest neighbouring sites have also been shown to experience an artificial gauge field in a graphene quantum simulator \cite{Lembessis+15}, while a trapping potential comprising a flat intensity profile and an inverse square power-law phase has been proposed for investigations on sonic horizons and artificial black holes \cite{Butera+16}. A Gaussian line with a phase gradient across it can be used to trap particles in optical tweezers, but at the same time cause them to flow \cite{Roichman+08}. As a test of our method's versatility, we have also chosen the more arbitrary patterns of a chicken and eggs \cite{Chicken} which have uncorrelated intensity and phase patterns.

The main metric for accuracy is the fidelity, which is defined as $F = \left| \sum_{n,m} \tau_{n,m}^{*} E^{\text{out}}_{n,m}\right|^{2}$ \cite{Goorden+14} and is evaluated over non-zero amplitude within the measure region. The light efficiency ($\eta$) is the fraction of light in the output plane that is in the region of interest.  A relative phase error $\epsilon_{\Phi}$ within the measure region and the non-uniformity error $\epsilon_{nu}$ for regions in the patterns that have a flat intensity \cite{Wu+15} are defined as:
\begin{align}
\epsilon_{\Phi} &= \frac{\sum_{n, m}|\left(\Phi_{n,m} - \varphi_{n,m} + P\right)|^{2}}{\sum_{n,m}|\Phi_{n,m}|^{2}},
\label{equation:phase_error} \\
\epsilon_{nu} &= \frac{\sum_{n, m}|M_{n,m}\left(\tilde{I}_{n,m} - I_{a}\right)|^{2}}{\sum_{n,m}|M_{n,m}\tilde{T}_{n,m}|^{2}},
\label{equation:nu_error}
\end{align}
where $P$ is a correction term to account for the cyclical nature of the phase, $M_{n,m}$ is a binary mask which is equal to one where the target intensity is approximately uniform and zero everywhere else and $I_{a} = \left(1/N\right)\sum_{n,m}M_{n,m}\tilde{I}_{n,m}$ is the average output intensity in the uniform region ($N$ is the total number of pixels in the measure region).

\begin{figure}[htb!]
\begin{center}
\includegraphics[width = \linewidth]{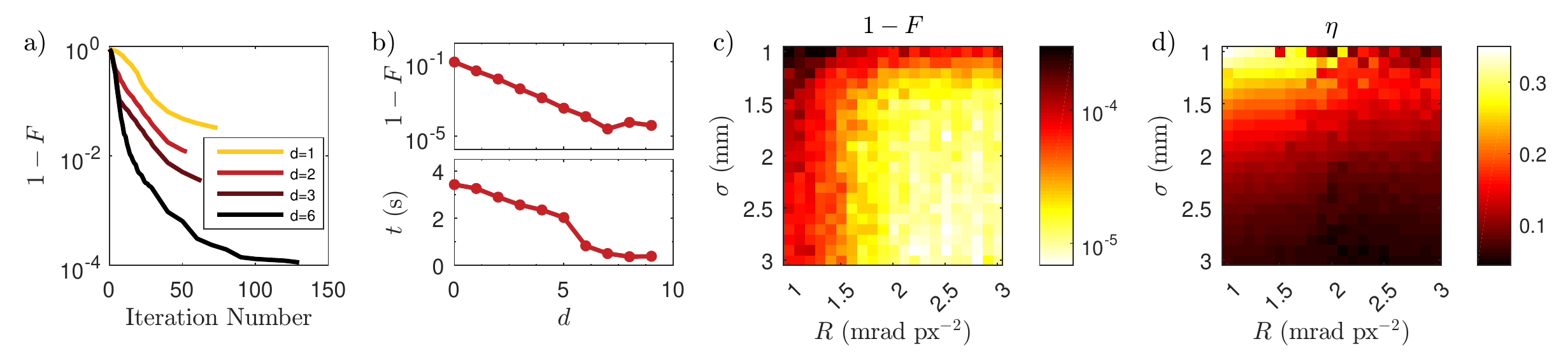}
\caption{a) Evolution of fidelity $F$ for the Gaussian Line pattern shown in Figure \ref{fig:results1}f) with $\sigma=1.5~\text{mm}$ and $R=2.3~\text{mrad~px}^{-2}$. At low values of the steepness $d$ of the cost function, the algorithm stagnates earlier and returns a lower fidelity hologram. b) The final fidelity and the time per iteration $t$ as a function of $d$. c) Fidelity and d) efficiency $\eta$ as a function of incident laser beam size $\sigma$ and quadratic guess phase curvature $R$. \label{fig:beam&quad}}
\end{center}
\end{figure} 

\begin{table}[htb!]
%\begin{center}
\centering
\caption{Error metrics for the calculated patterns in Figure \ref{fig:results1}, with optimal values of $\sigma$, $R$ and region of interest diameter ROI. \label{tabular:results1}}
  %\setlength{\tabcolsep}{10pt}
  %\scalebox{0.6}{
  \begin{tabular}{ | l | c c c | c c c c|}
  \hline
  Pattern & $\sigma$ & $R$ & ROI & $1-F$ & $\eta$ & $\epsilon_\Phi$ & $\epsilon_{nu}$\\
  & mm & mrad px$^{-2}$ & px & & \% & \% & \% \\ \hline
  a) Laguerre Gauss & 1.0 & 4.5 & 42 & ~$3.0\times 10^{-6}$~ & 41.5 & 0.0003 & 0.005 \\ %\hline
  b) Square Lattice & 1.2 & 4.5 & 124 & $1.6\times 10^{-5}$ & 10.6 & ~0.00009~ & 0.02 \\ %\hline
  c) Ring Lattice & 1.2 & 3.9 & 71 & $1.5\times 10^{-6}$ & 24.6 & 0.00006 & 0.001 \\ %\hline
  d) Graphene & 1.4 & 2.7 & 78 & $4.4\times 10^{-4}$ & 13.1 & 0.0003 & 0.010~ \\ %\hline
  e) Flat Top & 1.0 & 4.5 & 63 & $1.8\times 10^{-4}$ & 11.3 & 0.2 & 0.007 \\ %\hline
  f) Gaussian Line & 1.4 & 2.9 & 45 & $1.4\times 10^{-5}$ & 20.4 & 0.01 & 0.002 \\ %\hline
  g) Chicken \& Egg & 1.6 & 4.5 & 128 & $7.1\times 10^{-2}$ & 2.0 & 1.3 & - \\ \hline
  \end{tabular}
  %}
%\end{center}
\end{table}

For the example of the Gaussian line pattern (with $\sigma=1.5~\text{mm}$ and $R=2.3\times 10^{-3}~\text{mrad~px}^{-2}$) Figure \ref{fig:beam&quad}a) shows the evolution of the fidelity through the calculation for different values of the steepness parameter $d$ in Equation (\ref{equation:cost1_2}). Lower values of $d$ cause early stagnation of the algorithm into poor quality local minima. The maximum iteration number was reached for $d>6$, whilst the fidelity would increase at approximately the same rate for $d>4$ (only $d = 1,2,3$ and $6$ are shown in Figure \ref{fig:beam&quad} for clarity). It was found that a steeper cost function would not only lead to improved fidelities in the patterns, but also faster calculation times per iteration $t$ (Figure \ref{fig:beam&quad}b)). A typical minimization routine converges in $<200$ iterations at a total duration of $<75\text{s}$ with a standard desktop computer ($2.5~\text{GHz}$ processor). For all patterns shown in this article, we have used $d=9$.

For each pattern we perform an optimisation over the initialization conditions $\sigma$ and $R$ (see Figure \ref{fig:beam&quad}c)-d)). It was found that smaller incident laser beam sizes and reduced curvature in the guess phase led to higher light efficiency at a reduced fidelity. The beam size and curvature for the patterns in Figure \ref{fig:results1} were chosen to provide both good light efficiency whilst maintaining a high fidelity. The optimal values of calculated holograms are shown in Table \ref{tabular:results1}.

The authors of \cite{Wu+15} recently developed an IFTA for full-plane control of amplitude and phase, which they compared to a previous regionally-constrained algorithm \cite{Tao+15}. They find that the regionally-constrained algorithm is more accurate at the cost of light-utilisation efficiency, which has also been seen in amplitude-only control algorithms \cite{Pasienski+08,Harte+14} and in the present work. For far-field holograms of lines of continuous intensity with phase gradients, the regional algorithm gives $\epsilon_{nu} = 0.04\%$, $\epsilon_{\Phi}=1.63\%$ and $\eta = 3.48\%$, while the full-plane IFTA is less accurate ($\epsilon_{nu} = 3.48\%$ and $\epsilon_{\Phi}=3.77\%$) but achieves higher efficiency ($\eta = 77.84\%$). For our chosen cost function in Equation (\ref{equation:cost1_2}), the comparable continuous patterns amongst our range of targets (i.e. the Gaussian Line and Flat Top) are significantly smoother: we find $\epsilon_{nu}$ is lower by a factor 6-20 and $\epsilon_{\Phi}$ is lower by one or two orders of magnitude than the regional IFTA. The light-utilisation of the conjugate gradient optimised patterns is a factor 3-11 times higher than the regional IFTA, but between $15$-$53\%$ of the full-plane IFTA. We note that the freedom in choice of the cost function terms and their relative weightings could be exploited to prioritise the efficiency of light usage at the expense of accuracy or smoothness if this is of greater importance to a particular application.

\section{Experimental Verification}

\begin{figure}[ht!]
\begin{center}
\includegraphics[width=0.7\linewidth]{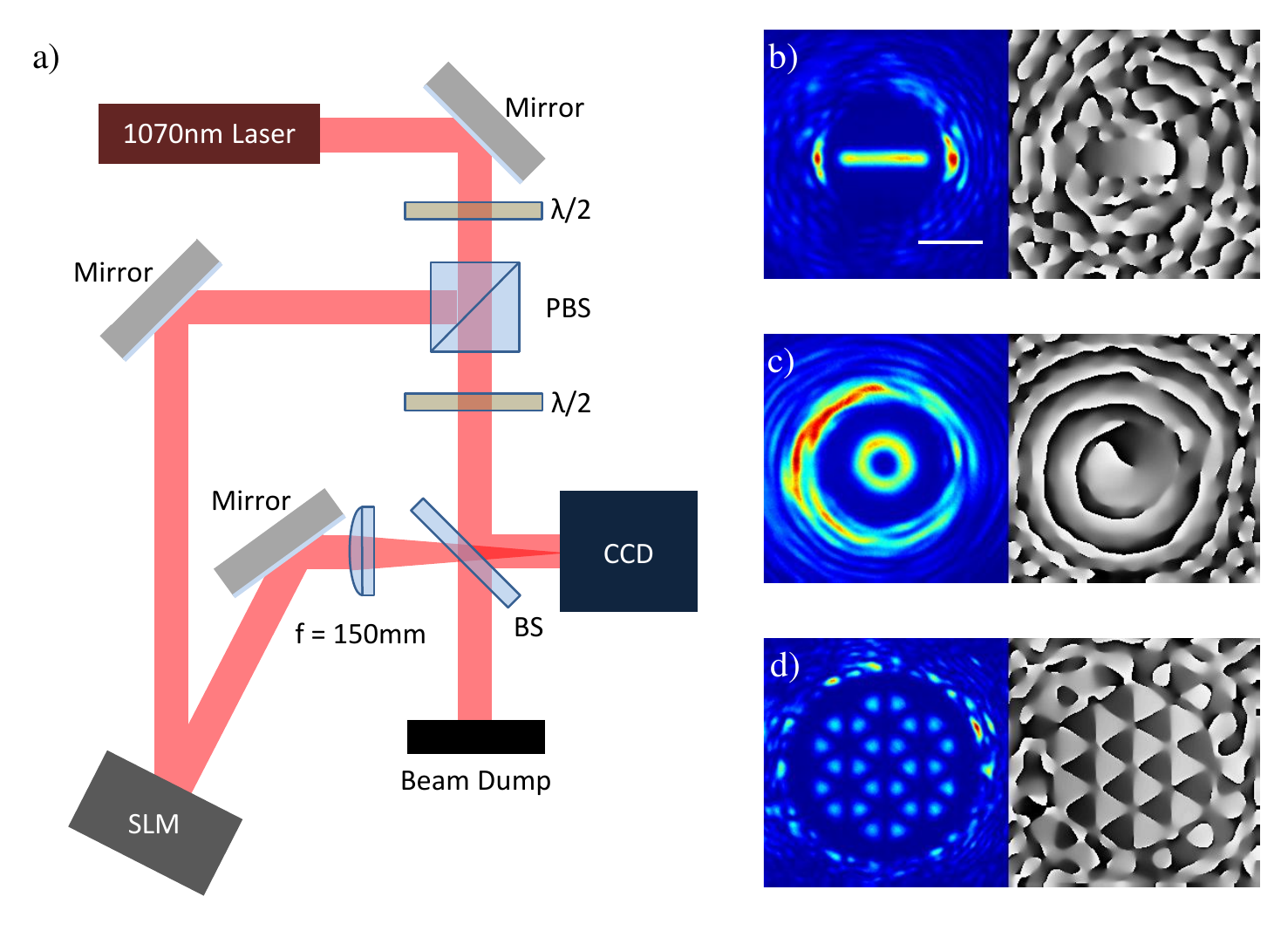}
\caption{a) Experimental Setup. The first $\lambda/2$ waveplate is used to vary the power between the reference and SLM beam, whilst the second waveplate is used to reorientate the polarisation of the reference beam to match the SLM beam for interference. b) - d) Measured intensity (left) and phase (right). The white scalebar in b) denotes $300~\mu\text{m}$, and is common to all images. Color scaling as in Figure \ref{fig:results1}. \label{fig:setup}}
\end{center}
\end{figure}

We verify the calculated holograms experimentally using the setup shown in Figure \ref{fig:setup}a). The output of a $1070~\text{nm}$ fiber laser (IPG YLP-5-1070-LP) is expanded to an experimentally-convenient $1/e^{2}$ waist of $3.0~\text{mm}$ and split using a polarising beam splitter. One path is phase-modulated as it is reflected ($14^{\circ}$ AOI) by a liquid crystal SLM (BNS P1920) and focussed onto a CCD camera (Thorlabs DCU200 Series) using an $f=150~\text{mm}$ achromatic doublet. The other path gives a reference beam which is optionally recombined with the modulated beam after the focussing optic to produce interference fringes which are used to extract the phase of the modulated light via the Fourier transform fringe analysis method \cite{Takeda+82}.

As shown in Figure \ref{fig:setup}b)-d) and detailed in Table \ref{tabular:results2}, the measured fidelities are lower than the numerical predictions, but could be improved by the addition of feedback \cite{Bruce+11,Bruce+15} or the characterisation of wavefront aberration in the optical system \cite{Cizmar+10,Zupancic+16}. We include the rescaled efficiency $\eta^{*}$, which is $\eta/0.45$, due to the $45\%$ diffraction efficiency of the SLM.  Higher diffraction efficiencies could be obtained by replacing the SLM with a micro-fabricated diffractive optical element.

\begin{table}[htb!]
\centering
%\begin{center}
\caption{Error metrics for the measured patterns in Figure \ref{fig:setup}. \label{tabular:results2}}
%  \setlength{\tabcolsep}{10pt}
 % \scalebox{0.6}{
  \begin{tabular}{ | l | c c c c | c c c c |}
  \hline
  & \multicolumn{4}{c|}{Theory} & \multicolumn{4}{c|}{Experiment} \\
  Pattern & $F$ & $\eta$ & $\epsilon_{\Phi}$ & $\epsilon_{nu}$ & $F$ & $\eta^{*}$ & $\epsilon_{\Phi}$ & $\epsilon_{nu}$ \\
  & & \% & \% & \% & & \% & \% & \% \\ \hline
  Gaussian Line & 0.99996 & 8.3 & 0.005 & 0.004 & 0.97 & 7.8 & 2.76 & 0.48 \\ %\hline
  % 97.4\%
  Laguerre Gauss & 0.99999 & 8.4 & 0.0004 & 0.004 & 0.97 & 7.8 & 2.59 & 0.52 \\ %\hline
  % 96.5\%
  Graphene & 0.9996 & 7.0 & 0.0004 & 0.015 & 0.96 & 6.2 & 1.85 & 0.42 \\ \hline
  % 95.6\%
  \end{tabular}
%  }
%\end{center}
\end{table}

\section{Conclusion}

We have demonstrated that smooth, high fidelity light patterns with independent control over the amplitude and phase can be generated with a single phase-only SLM. The holograms calculated with the conjugate gradient minimisation approach surpass the accuracy and smoothness of previous IFTA approaches. We note that our approach achieves comparable results in $F$ and $\eta$ for image-quality holograms to the super-pixel method for DMDs\cite{Goorden+14}, and improved $F$ for the LG mode, at the expense of constraining the pattern to a subset of the output plane. 

This approach to hologram calculation is compatible with existing methods for the generation of multi-wavelength holographic optical traps \cite{Bowman+15}. In this work we have concentrated on using a fast Fourier transform as the propagator $\mathcal{P}$. However, we find that near-field patterns calculated using Angular Spectrum Wavefront Propagation \cite{Goodman+96} achieve comparable fidelity, efficiency and smoothness. The accurate control over amplitude and phase will be crucial to a future research direction in the design of axially-structured light fields.

%\section{Conclusions}

\section*{Funding}
Leverhulme Trust (RPG-2013-074); EPSRC (EP/G03673X/1; EP/L015110/1).

\section*{Acknowledgments}
We thank 
% Markus Greiner
L. Walker and T. Doherty for useful discussions and T. Scrivener and P. Collins for the loan of the SLM. 

%%%%%%%%%%%%%%%%%%%%%%% References %%%%%%%%%%%%%%%%%%%%%%%%%

\end{document}